\documentclass[a4paper]{jpconf}
\usepackage{graphicx}
\begin{document}
\title{Coexistence of $d$-wave superconductivity and antiferromagnetism induced by paramagnetic depairing}

\author{Yuhki Hatakeyama and Ryusuke Ikeda}

\address{Department of Physics, Kyoto University, Kyoto 606-8502, Japan}

\ead{ryusuke@scphys.kyoto-u.ac.jp}


\begin{abstract}
It is shown theoretically that, in the superconducting state with $d_{x^2-y^2}$-pairing, a strong Pauli paramagnetic depairing (PD) induces not only the modulated Fulde-Ferrell-Larkin-Ovchinnikov (FFLO) superconducting state but also an incommensurate antiferromagnetic (AFM) (or spin-density wave) ordering with ${\bf Q}$-vector nearly parallel to a gap node. In this mechanism of field-induced coexistence of the $d$-wave superconducting and AFM orders, a pair-density wave does not have to be assumed. It is argued that this is the common origin of {\it both} the coexistent FFLO and AFM phases of CeCoIn$_5$ {\it and} the AFM quantum critical behavior around the superconducting $H_{c2}(0)$ seen in several unconventional superconductors. 
\end{abstract}

\section{Introduction}
At present, we have two intriguing issues on a magnetic order or fluctuation occurring close to $H_{c2}(0)$ in $d$-wave paired superconducting states. One is the antiferromagnetic (AFM) quantum critical behavior reflected in transport measurements around $H_{c2}(0)$ of heavy fermion superconductors CeCoIn$_5$ [1-3], pressured CeRhIn$_5$ [4], NpPd$_5$Al$_2$ [5], and Tl-based cuprates [6]. The other is the AFM order [7] in the high field and low temperature (HFLT) phase [8] of CeCoIn$_5$ which has been identified, based on measurements [9,10] and theoretical explanations [11,12] of elastic properties and doping effects, with a Fulde-Ferrell-Larkin-Ovchinnikov (FFLO) state [8,13]. It is notable that the materials listed above have a couple of common features, such as the $d$-wave pairing and a strong Pauli paramagnetic depairing (PD). 

In this report, we point out that, in {\it nodal} $d$-wave superconductors, a field-induced enhancement of PD tends to induce an AFM ordering in the superconducting phase just below $H_{c2}(0)$. Conventionally, the AFM ordering is suppressed by the superconducting ordering in zero field [14], and, even in nonzero fields, the quasiparticle damping effect brought by the AFM fluctuation usually suppresses a relatively weak PD effect, suggesting a competition between the two orderings. It is found, however, that a strong PD rather favors coexistence of a $d$-wave superconductivity and an {\it incommensurate} AFM order, leading to an enhancement of AFM ordering or fluctuation just below $H_{c2}(0)$. 

\section{Model and Calculation}
We work in a BCS-like electronic Hamiltonian [14] with the superconducting energy gap $\Delta$ and the AFM moment ${\bf m}$. By treating $\Delta$ at the mean field level and $m \equiv |{\bf m}|$ as a fluctuation, respectively, the free energy describing the two possible orderings in zero field is described by 
\begin{eqnarray}
{\cal F}({\bf H} = 0) &=& \int d^3r \, g^{-1} |\Delta({\bf r})|^2 - T \, {\rm ln} \, {\rm Tr}_{c, c^\dagger, m} \exp[- (H_{\Delta \, m} - \mu N)/T], 
\nonumber \\
H_{\Delta \, m} - \mu N &=& \sum_{\bf q} \frac{1}{U} |{\bf m}({\bf q})|^2 + \sum_{{\bf k},\alpha,\beta} \, {\hat c}^\dagger_{{\bf k}, \alpha} \, \xi({\bf k}) \, \delta_{\alpha,\beta} \, {\hat c}_{{\bf k},\beta} \nonumber \\
&-& \sum_{{\bf q},\alpha,\beta} \biggl( \Delta({\bf q}) {\hat \Psi}^\dagger({\bf q}) + m_{\nu}({\bf q}) \, {\hat S}^\dagger_{\nu}({\bf q}) + {\rm h.c.} 
\biggr), 
\end{eqnarray}
where ${\hat \Psi}({\bf q}) = - {\rm i} (\sigma_y)_{\alpha,\beta} \, \sum_{\bf k} w_{\bf k} \, {\hat c}_{-{\bf k}+{\bf q}/2, \alpha} {\hat c}_{{\bf k}+{\bf q}/2, \beta}/2$, ${\hat S}_{\nu}({\bf q})= (\sigma_\nu)_{\alpha,\beta} \, \sum_{\bf k} {\hat c}^\dagger_{{\bf k}-{\bf q}, \alpha} {\hat c}_{{\bf k}+{\bf Q}, \beta}/2$, ${\hat c}^\dagger_{{\bf k}, \alpha}$ creates a quasiparticle with spin index $\alpha$ and momentum ${\bf k}$, $\sigma_\nu$ are the Pauli matrices, $\mu$ is the chemical potential, and the positive parameters $g$ and $U$ are the attractive and repulsive interaction strengths leading to the superconducting and AFM orderings, respectively. The gap function $w_{\bf k}$ satisfies $w_{{\bf k}+{\bf Q}} = - w_{\bf k}$ for the $d_{x^2-y^2}$-pairing state, 
and the dispersion $\xi({\bf k})$ satisfies $\xi({\bf k}) = - \xi({\bf k}+{\bf Q}) + T_c \delta_{\rm IC}$, where ${\bf Q}$ is the {\it commensurate} AFM modulation wavevector and is $(\pi$, $\pi)$ for the $d_{x^2-y^2}$-pairing. A small deviation from the perfect nesting is measured by a small parameter $\delta_{\rm IC}$ for a nearly-free electron model. If the tight-binding model with dispersion $\xi({\bf k}) = -t_1 ({\rm cos}(k_x a) + {\rm cos}(k_y a)) - t_2 {\rm cos}(k_x a) {\rm cos}(k_y a) - \mu$ and the lattice constant $a$ is directly used in examining an AFM ordering, the corresponding incommensurability is measured by the second term of the above $\xi({\bf k})$. In a nonzero field (${\bf H} \neq 0$) of our interest, the Zeeman term $\gamma_B H (\sigma_z)_{\alpha,\beta}$ needs to be added to $\xi({\bf k}) \delta_{\alpha,\beta}$. At least in the case with a continuous $H_{c2}$-transition like Fig.2 (a) below, the orbital depairing needs to be incorporated through the familiar quasiclassical treatment on the quasiparticle Green's function [13]. 

\begin{figure}[b]
\scalebox{1.4}[1.4]{\includegraphics{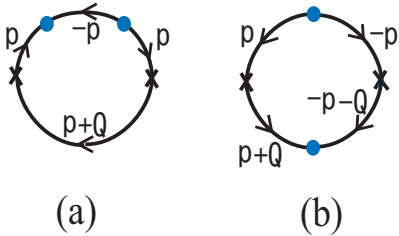}}
\caption{Diagrams describing (a) $\chi^{(n)}$ and (b) $\chi^{(an)}$ up to O($|\Delta|^2$), where the cross denotes the particle-hole vertex on the AFM fluctuation, while the filled circle implies the particle-particle vertex on $\Delta$ or $\Delta^*$. }
\label{fig.1}
\end{figure}
To see the position of the AFM ordering, it is convenient to examine the Gaussian AFM fluctuation term ${\cal F}_{m}$ in the free energy ${\cal F}$, ${\cal F}_m = T \sum_{\Omega} {\rm ln} \, {\rm det}[ \, U^{-1} \delta_{{\bf q}, {\bf q}'} - \chi_{{\bf q}, {\bf q}'}(\Omega) \, ]$, where 
\begin{eqnarray}
\chi_{{\bf q}, {\bf q}'}(\Omega) &=& \! \int_0^{T^{-1}} d\tau \, \langle \, T_\tau \, {\hat S}^\dagger_\nu({\bf q};\tau) \, {\hat S}_\nu({\bf q}';0) \, \rangle \, e^{i \Omega \tau}, 
\end{eqnarray}
and ${\hat S}_\nu({\bf q};\tau)$ denotes ${\hat S}_\nu({\bf q})$ at imaginary time $\tau$. 
For the moment, we focus on the Pauli limit with no orbital depairing and with uniform $\Delta$ in which $\chi_{{\bf q}, {\bf q}'}(\Omega) = [\chi^{({\rm n})}({\bf q}, \Omega) + \chi^{({\rm an})}({\bf q}, \Omega)] \delta_{{\bf q}, {\bf q}'}$, and ${\cal F}_{m} = - T \sum_{{\bf q}, \Omega} {\rm ln} X({\bf q}, \Omega)$, where $X^{-1}({\bf q}, \Omega) = U^{-1} - \chi^{({\rm n})}({\bf q}, \Omega) - \chi^{({\rm an})}({\bf q}, \Omega)$. A second order AFM ordering occurs when $X_0^{-1} = X^{-1}(0,0)=0$. The O($|\Delta|^2$) terms in $\chi^{({\rm n})}$ and $\chi^{({\rm an})}$ are expressed by Fig.1 (a) and (b), respectively. They have been studied previously [14] in $H=0$ case, where $\chi_s(\Delta) \equiv \chi^{({\rm n})}(0,0) - \chi^{({\rm n})}(0,0)|_{\Delta=0} + \chi^{({\rm an})}(0,0)$ taking the form 
\begin{eqnarray}
\chi_s(\Delta) &=& T \int \! \frac{d^3p}{(2 \pi)^3} \sum_{\varepsilon, \sigma} \biggl[ 2 w_{\bf p}^2 \, ({\cal G}_{\varepsilon, \sigma}({\bf p}))^2 \, \Delta^* {\cal G}_{- \varepsilon, -\sigma}(-{\bf p}) \, \Delta \, {\cal G}_{\varepsilon, {\overline \sigma}}({\bf p} + {\bf Q}) \nonumber \\
&-& w_{\bf p} w_{{\bf p}+{\bf Q}} \, {\cal G}_{\varepsilon, \sigma}({\bf p}) \, \Delta^* {\cal G}_{- \varepsilon, -\sigma}(-{\bf p}) {\cal G}_{\varepsilon, {\overline \sigma}} ({\bf p} + {\bf Q}) \Delta {\cal G}_{-\varepsilon, -{\overline \sigma}}(-{\bf p} - {\bf Q}) \biggr]
\end{eqnarray}
behaves like $T^{-2}$ in $T \to 0$ limit and is negative so that the AFM ordering is suppressed by superconductivity [14]. In eq.(3), ${\cal G}_{\varepsilon, \sigma}({\bf p}) = ({\rm i}\varepsilon - \xi_{\bf p} + \gamma_B H \sigma)^{-1}$ denotes the quasiparticle Green's function defined in the normal state. The last term implies $\chi^{({\rm an})}(0,0)$. 

To explain effects of strong PD, let us first explain the ${\bf m} \parallel {\bf H}$ case in which ${\overline \sigma}=\sigma$. In this case, the two terms in eq.(3) are found to take the same form as the coefficient of the O($|\Delta|^4$) term of the superconducting Ginzburg-Landau (GL) free energy and thus, change their sign upon cooling [13]. Hence, $\chi_s(\Delta)$ becomes positive for stronger PD, leading to a lower ${\cal F}_m$, i.e., an {\it enhancement} of the AFM ordering in the superconducting phase. As well as the corresponding PD-induced sign-change of the O($|\Delta|^4$) term which leads to the first order $H_{c2}$-transition [13], the PD-induced positive $\chi_s$ is also unaffected by inclusion of the orbital depairing. 

In ${\bf m} \perp {\bf H}$ where ${\overline \sigma}= - \sigma$, a different type of PD-induced AFM ordering occurs in a $d$-wave pairing case with a gap node along ${\bf Q}$ where $w_{{\bf p}+{\bf Q}} = - w_{\bf p}$ : In this case, the first term of eq.(3) arising from $\chi^{({\rm n})}(0,0)$ remains negative as in zero field case and becomes $-N(0)|\Delta|^2/[2 (\gamma_{\rm B} H)^2 ]$ in $T \to 0$ limit with no PD-induced sign change, where $N(0)$ is the normal density of states. Instead, the last term of eq.(3) implying $\chi^{({\rm an})}(0,0)$ and thus, $\chi_s$ are divergent like $N(0) [ \, |\Delta|/(\gamma_{\rm B} H) \,]^2 \, |{\rm ln}[{\rm Max}(t, |\delta_{\rm IC}|)]|$ in $T \to 0$ limit while keeping their positive signs owing to the relation $w_{{\bf p}+{\bf Q}}\cdot w_{\bf p} < 0$, where $t=T/T_c$. This divergence is {\it unaffected} by including the orbital depairing. That is, in the $d_{x^2-y^2}$-wave case with ${\bf Q}=$ ($\pi$, $\pi$), the AFM order tends to occur upon cooling in ${\bf m} \perp {\bf H}$. In contrast, $\chi^{({\rm an})}(0,0)$ is also negative in the $d_{xy}$-wave case with the same ${\bf Q}$ satisfying $w_{\bf p} w_{{\bf p}+{\bf Q}} > 0$ so that the AFM ordering is suppressed with increasing $H$. 

\section{Examples of Phase Diagrams}

\begin{figure}[t]
\scalebox{0.7}[0.7]{\includegraphics{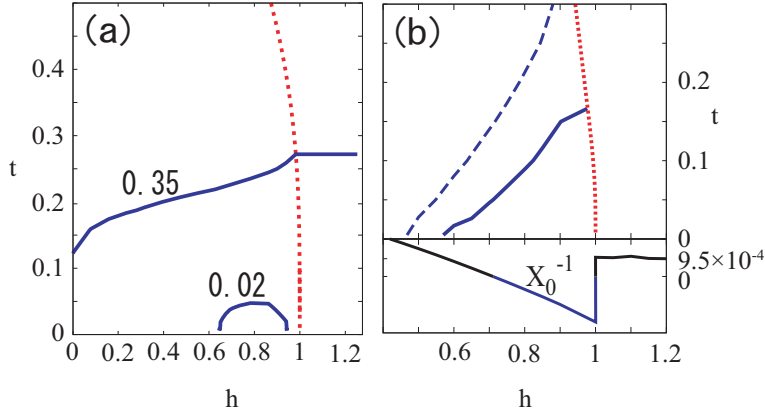}}
\caption{Typical $t$ ($=T/T_c$) v.s. $h=H/H_{c2}(0)$ phase diagrams (a) following from the use of $\alpha=0.3$ leading to a second order $H_{c2}$-transition even in low $t$ limit and (b) in the Pauli limit with a first order $H_{c2}$-transition in low $t$, respectively. In both figures, an AFM phase can occur below a solid curve on which $X^{-1}_0=0$, and each nearly vertical dotted curve is the corresponding $H_{c2}(T)$-curve. Note that, in Fig.2(a), two AFM phase boundaries for $T_{\rm N}/T_c=0.35$ and $0.02$ are shown in a single figure. Figure 2 (b) was obtained from the corresponding calculation in the tight binding model with the parameters $U=33$, $t_1=100$, and $t_2=0.25$ in the unit of $T_c$ and by taking account of the {\it full} $\Delta$ dependence with no limitation to the O($|\Delta|^2$) term. The dashed curve denotes the possible upper limit of the AFM transition temperatures at which $\chi_s=0$. The lower panel of (b) is the $h$-dependence of $X^{-1}_0$ at $t=0.05$. }
\label{fig.2}
\end{figure}

In this section, examples of the resulting low temperature phase diagram near $H_{c2}(0)$ will be presented. In the BCS-like model (1) and up to the O($|\Delta|^2$) terms (see Fig.1), the $H_{c2}$-transition, i.e., the mean field superconducting transition in $H \neq 0$, is of second order even at lower temperatures for $\alpha \equiv \gamma_B H_{\rm orb}(0)/(2 \pi T_c) \leq 0.3$ (see Fig.2(a)), where $\alpha$ is nothing but the Maki parameter except a difference in the numerical factor, while it becomes of first order for larger $\alpha \simeq 1.1$ [13]. It is reasonable to expect the former to correspond to the case of CeRhIn$_5$ under a pressure [4]. Figure 2(a) is one of the phase diagrams in such a case, where the Neel temperature $T_{\rm N}$ in the normal state with perfect nesting or $U$ was assumed to be the only parameter measuring the pressure dependence. The actual AFM transition temperature in $H > H_{c2}(0)$ for $T_{\rm N}/T_c=0.02$ and $0.35$ are zero and less than $0.35 T_c$, respectively, because of the finite $\delta_{\rm IC} \simeq 0.6$ used in the calculation. Reflecting the AFM ordering enhanced by PD, the decrease of $T_{\rm N}$, corresponding to an increase of pressure, results in the shrinkage of the AFM phase just below $H_{c2}(0)$, which reduces to an apparent AFM quantum critical point by a further increase of pressure. 

Figure 2(b) is the corresponding result in the tight binding model in the Pauli limit with no orbital depairing (vortices). The $H_{c2}$-transition is of first order in the temperature range shown there. Due to the discontinuous nature of the $H_{c2}$-transition, an apparent AFM quantum critical point is estimated, in $h = H/H_{c2}(0) > 1$, to lie at a lower field than $H_{c2}(0)$ in spite of the PD-induced AFM ordering just below $H_{c2}(0)$. This is consistent with the observations in CeCoIn$_5$ [2,7]. We note that the anomalous doping effect in CeCoIn$_5$ [10] cannot be explained without a spatial modulation of $|\Delta|$ in the HFLT phase [12], implying that both the AFM and FFLO orders coexist in the HFLT phase of CeCoIn$_5$. Calculation results in the case including the FFLO structure will be reported elsewhere [15]. 

We note that, in the present theory explaining the AFM order just below $H_{c2}(0)$ in CeCoIn$_5$ with strong PD, the assumption [16] of an additional pairing channel (pair-density wave) is unnecessary, and that both of the AFM order [7] and other observations, such as the anomalous doping effect [10], in the HFLT phase of CeCoIn$_5$ are explained consistently if the FFLO modulation in the HFLT phase is assumed.  

This work was financially supported by Grant-in-Aid for Scientific Research [No. 20102008 and 21540360] from MEXT, Japan. 




\end{document}